\RequirePackage{fix-cm}
\documentclass[nonatbib]{svjour3}
\pdfoutput=1
\smartqed  
\makeatletter
\let\c@author\relax
\makeatother
\usepackage[backend=bibtex,style=nature,url=false,isbn=false,hyperref=false ]{biblatex}
\usepackage{graphicx}
\usepackage{mathptmx}      
\usepackage{latexsym}
\usepackage{lineno}
\usepackage{textcomp}
\usepackage{color}
\usepackage{xcolor}
\usepackage{setspace}
\usepackage{multirow}
\usepackage{url}
\journalname{Scientific Reports}
\begin{document}

\title{Alternative climatic steady states near the Permian-Triassic Boundary}


\author{C. Ragon$^1$ \and  C. V\'erard$^2$ \and J. Kasparian$^1$  \and M. Brunetti$^{1,*}$}


\institute{}

\date{Received: date / Accepted: date}

\maketitle

\smallskip\noindent 
$^1$ Group of Applied Physics and Institute for Environmental Sciences, University of Geneva, 66 Bd Carl-Vogt, CH-1211 Geneva 4, Switzerland

\smallskip\noindent 
$^2$ Section of Earth and Environmental Sciences, University of Geneva, 13 Rue des Mara\^ichers, CH-1205 Geneva, Switzerland 

\smallskip\noindent 
$^*$ Corresponding author: Maura Brunetti, maura.brunetti@unige.ch 

\bigskip

\begin{abstract}
Due to spatial scarcity and uncertainties in sediment data, initial and boundary conditions in deep-time climate simulations are not well constrained.
On the other hand, depending on these conditions, feedback mechanisms in the climate system compete and balance differently. This opens up the possibility to obtain multiple steady states in numerical experiments. 
Here, we use the MIT general circulation model 
to explore the existence of such alternative steady states around the Permian-Triassic Boundary (PTB). We construct the corresponding bifurcation diagram,  taking into account processes on a timescale of thousands of years, in order to identify the stability range of the steady states and tipping points as the atmospheric CO$_2$ content is varied.
We find three alternative steady states with a difference in global mean surface air temperature of about 10~$^\circ$C.
We also examine how these climatic steady states are modified when feedbacks operating on comparable or longer time scales are included, namely vegetation dynamics and air-sea carbon exchanges. Our findings on multistability provide a useful framework for explaining the climatic variations observed in the Early Triassic geological record, as well as some discrepancies between numerical simulations in the literature and geological data at PTB and its aftermath. 
\end{abstract}

\section{Introduction}
\label{section:intro}

Earth's climate is a complex system that results from the balance between a spatially inhomogeneous distribution of energy received from the Sun, and interactive and dissipative processes occurring at various temporal and spatial scales~\cite{GhilLucarini2020}.
A multitude of feedback mechanisms takes place in the system, some reducing (e.g., Stefan-Boltzmann radiative feedback) and others amplifying (e.g., ice-albedo feedback) the effect of an initial change in the average temperature. Depending on how these feedbacks compensate or interact with each other, the climate system can reach different steady states (or \textit{attractors}) under the same forcing, a phenomenon called `multistability'~\cite{GhilLucarini2020}.

Global means of state variables, like surface air temperature, change in a quasi-linear manner in response to a forcing when the attractor remains stable. However, abrupt 
climate changes can occur when the attractor loses its stability through different tipping mechanisms~\cite{ashwin2012,feudel2018}. In particular, a bifurcation-induced tipping (B-tipping) happens at the stability limit of an attractor, called tipping point. 
In this case, a shift from one attractor to another can be associated to an abrupt modification of climatic conditions under a small change in forcing developing on much longer timescales, such as variations of the astronomical cycles ($10^4$-$10^5$~yr)~\cite{crucifix2006} or tectonic movements ($10^6$~yr) \cite{raymo1992}. Noise-induced tipping (N-tipping) develops when the variability of the dynamics on the attractor exceeds the height of the barrier separating two basins of attraction.  This occurs because of the amplification of specific feedback mechanisms that can be triggered, for example, by volcanic activity~\cite{baum2022} acting on timescales of $10^5$-$10^6$~yr in large igneous provinces, or by the biological pump 
under a spatial shift of biome types~\cite{elke2020}. 
Also shock-induced tipping (S-tipping) can be caused by volcanic activity, or alternatively by asteroid impacts~\cite{impact1989}, or any other mechanism inducing a shift to another attractor on a shorter time scale than N-tipping. Finally,  forcing that varies on a time scale faster than the relaxation time towards the attractor can give rise to rate-induced tipping (R-tipping), even in the absence of multistability~\cite{feudel2018}.

Present anthropogenic CO$_2$ emissions are pushing the system towards critical thresholds~\cite{armstrong2022}.
Thus, a better understanding of these tipping mechanisms becomes essential through, for example, the study of the evolution of Earth's climate in the past~\cite{esd-15-41-2024}. 
Signatures of global climatic transitions are indeed found in paleoclimate proxy records for several periods of  Earth's history~\cite{boers2022}
such as during the Snowball Earth episodes in the Neoproterozoic era~\cite{Pierrehumbert2005,Hoffman2017},
the Eocene-Oligocene transition~\cite{Hutchinson2021}, 
the glacial-interglacial cycles~\cite{ferreira2018},
and the whole Cenozoic~\cite{rousseau2023}

Here, we are interested in
the climatic oscillations observed just after the Permian-Triassic Boundary (PTB) mass extinction ($\sim 252$~Ma), the most severe of the Phanerozoic~\cite{raup1979,macleod2014,stanley2016}. 
As a consequence of the  volcanic activity of the Siberian Traps~\cite{campbell1992,REICHOW20099}
the global carbon cycle entered a perturbed state that persisted for nearly 5~Myr in the Early Triassic and beyond, until a new equilibrium state was reached in the Anisian~\cite{sun2012,romano2013,goudemand2019,leu2019,widmann2020}. 
The observed fluctuations in the carbon isotope record~\cite{payne2004}
along with successive 
diversification-extinction cycles of the
nekton~\cite{leu2019}
and ecological crises of terrestrial plants~\cite{HERMANN2011,elke2020} are all indicative of the climatic changes that occurred in the Early Triassic, with variations in global temperature on the order of 
10~$^\circ$C from the thermal maximum in the late Smithian to cold climates in early Spathian times~\cite{widmann2020}.

However, despite the significant  efforts of the scientific community to reconstruct the climatic oscillations following the PTB, large uncertainties remain in the timing and causal relationships. This implies that numerical modelling needs to consider a wide range of initial and boundary conditions for simulating this geological interval. In this context, the framework of multistability, where an ensemble of initial conditions is explored to find the possible attractors under the same forcing and boundary conditions, and the subsequent construction of the so-called bifurcation diagram (BD) where the forcing is varied~\cite{brunetti2023}, seems particularly useful for identifying possible scenarios of tipping mechanisms.


Multistability has been observed in the entire hierarchy of climate models, from energy balance models to general circulation models with slab or dynamical ocean using aquaplanet or idealized continental configurations~(see \cite{brunetti2019, GhilLucarini2020} and references therein).
When performing numerical simulations, only selected spatial and temporal scales can be modelled due to the unavoidable compromise between the targeted phenomena under study and computational costs. This is why only a limited number of feedbacks can in general be dynamically included in numerical simulations, depending on the chosen spatial grid resolution and the model complexity. 
However, less computationally expensive numerical techniques can be used, like asynchronous or `offline' coupling~\cite{liu1999} 
or more detailed descriptions that are only activated in the last part of the simulations.

Here, we consider the Permian-Triassic paleogeography obtained with PANALESIS~\cite{verard2019panalesis,verard2021}, which offers the advantage of providing a plate tectonic reconstruction with full
seabed bathymetry as well as land topography.
We explore the existence of multiple attractors using the MIT general circulation model~\cite{marshall_finite-volume_1997,marshall_hydrostatic_1997,adcroft_implementation_2004}
in a fully coupled configuration that includes atmosphere, ocean, thermodynamic sea ice, land and fixed paleogeography, as described in the Methods section. 
We systematically consider an ensemble of initial conditions under fixed boundary conditions and external forcing to find multiple attractors in Section~\ref{section:attractors}. 
Then, 
by varying the atmospheric CO$_2$ content, we construct the BD in Section~\ref{section:BD}. 
Since the longest dynamical timescale considered in the model is the relaxation time of deep ocean (of the order of $10^3$~yr), we use the above mentioned numerical techniques to include feedbacks acting on a comparable timescale, namely 
the evolution of the vegetation cover and air-sea carbon exchanges, 
as described in Section~\ref{section:carboncycle}. 
We draw our conclusions and discuss further developments in Section~\ref{section:conclusions}.

\section{Results and discussion} 
\label{section:results}

\subsection{Description of three steady states}
\label{section:attractors}

Under the same forcing represented by an atmospheric pCO$_2$ of 320~ppm and boundary conditions provided by PANALESIS for the Permian-Triassic paleogeography, as specified in the Methods Section, the system relaxes over a time scale of  several thousand years towards one of three climatic attractors. 
Much colder attractors, such as the waterbelt or snowball states, are excluded from this study since they are not relevant for the Early Triassic climates. 
Here, we present the diagnostics calculated over the last simulated 30~yr for each attractor.  

\begin{figure*}[ht!]  
\includegraphics[width=\textwidth]{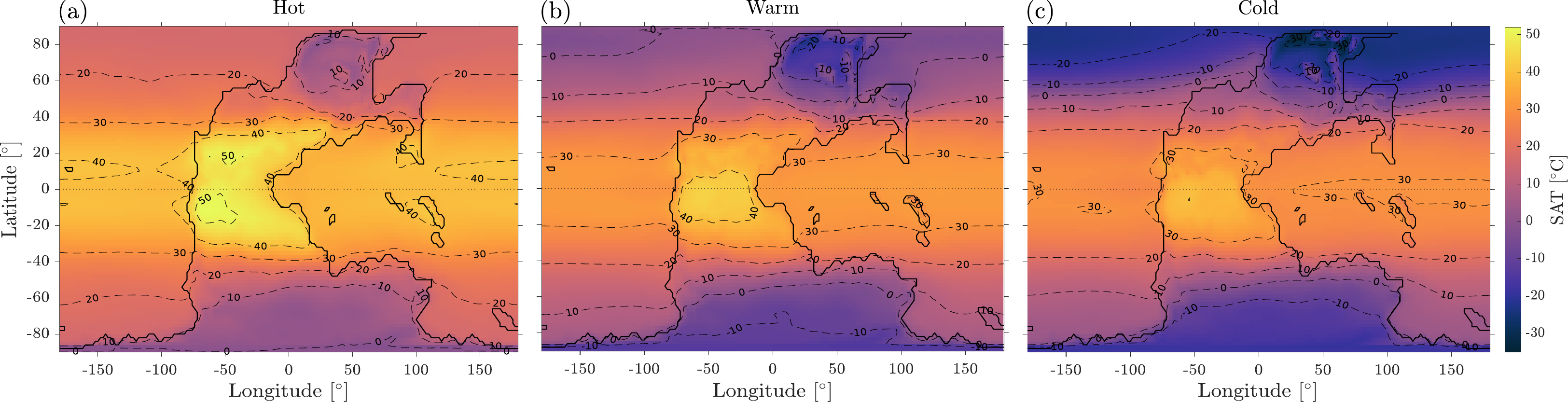}    
\includegraphics[width=\textwidth]{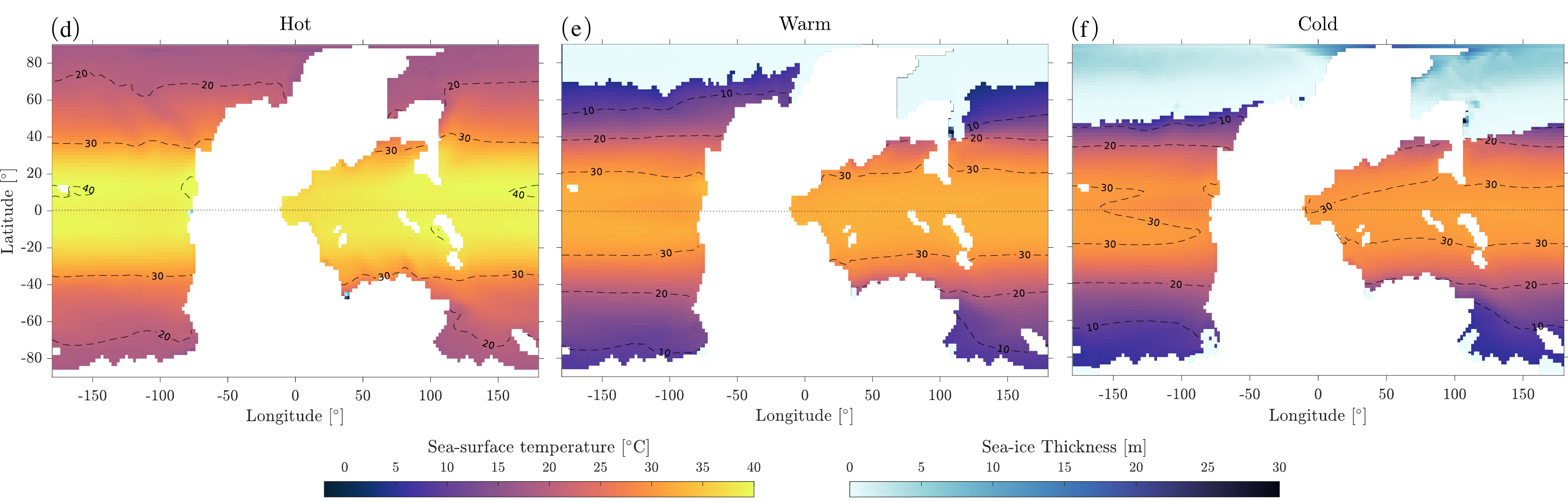}
    \caption{
    {\it Top row:} surface air temperature;
    {\it bottom row:} sea surface temperature and sea ice thickness, for the hot (a, d), warm (b, e), and cold (c, f) states. White area corresponds to land. 
    }
    \label{fig:sat}
\end{figure*}

Sea surface temperature, sea ice thickness and surface air temperature (SAT) are shown in Fig.~\ref{fig:sat}.
The global mean SAT ranges from 30.9~$^\circ$C for the {\it hot state} where no ice is present, to 21.56~$^\circ$C for the intermediate {\it warm state} where a small ice cap reaches $\sim 70^\circ$~N (locally 40$^\circ$~N to the East of Pangea), and to 17.20~$^\circ$C for the {\it cold state} with a perennial ice cap in the northern polar region down to $\sim 45^\circ$~N on average. A seasonal, small and thin sea-ice layer forms in the southern polar region in both the warm and cold states. 



\begin{table*}[t]
\caption{Global mean values averaged over the last simulated 30~yr  with associated standard deviation in parentheses derived from inter-annual variability, for each attractor. NH: northern hemisphere, SH: southern hemisphere.}
\label{tab:globmean}
\centering
\begin{tabular}{lccc}
\hline
 & Hot state & Warm state & Cold state \\ 
\hline
{\it Conservation diagnostics} & & & \\
TOA imbalance [W~m$^{-2}$] & $-0.1~(1)$ & $-0.4~(1)$ & $-0.3~(2)$\\
Ocean surface imbalance [W~m$^{-2}$] & $0.0~(2)$ & $-0.1~(2)$ & $0.07~(3)$\\
Water-mass budget [$10^{-5}$ g~m$^{-2}$~s$^{-1}$] & $0~(2)$ & $0~(2)$ & $-0~(2)$\\
Ocean temperature drift [$^\circ$C~century$^{-1}$] & $0.005~(1)$ & $-0.015~(1)$ & $0.006~(2)$\\
Salinity drift [10$^{-3}$ psu~century$^{-1}$] & $-7.85~(6)$ & $-6.76~(6)$ & $-5.7~(4)$ \\
\\
\hline
{\it Climatic variables} & & & \\
SAT [$^\circ$C] & $30.90~(7)$ & $21.56~(9)$ & $17.20~(9)$\\
Temperature gradient [$^\circ$C] & $18.3~(2)$ & $21.0~(2)$ & $24.3~(2)$\\
NH temperature gradient [$^\circ$C] & $17.2~(2)$ & $22.3~(2)$ & $28.5~(2)$\\
SH temperature gradient [$^\circ$C] & $19.4~(2)$ & $19.7~(3)$ & $20.0~(3)$\\
Ocean temperature [$^\circ$C] & $16.9987~(5)$ & $7.145~(2)$ & $5.289~(2)$\\
Precipitation [g~m$^{-2}$~s$^{-1}$] & $0.0477~(2)$ & $0.0377~(2)$ & $0.0352~(1)$\\
Evaporation [g~m$^{-2}$~s$^{-1}$] & $0.0477~(2)$ & $0.0377~(2)$ & $0.0352~(1)$ \\
Sea ice extent [10$^6$ km$^2$] & $0.008~(9)$ & $5.3~(9)$ & $34.3~(4)$\\
Sea ice thickness [m] & - & $0.079~(4)$ & $0.864~(5)$ \\
Latitude of sea ice boundary (NH) & - & $70$ & $45$\\
\hline
\end{tabular}
\end{table*}


Table~\ref{tab:globmean} shows global averages of selected key state variables and conservation diagnostics. The three attractors have a closed surface energy balance (i.e., lower than 0.1~W~m$^{-2}$ in absolute value, since we run the simulations towards equilibrium) over the ocean, which is the dominant component as it covers 69\% of the Earth surface at PTB. 
The TOA energy budget is also well closed, ranging from $-0.1$ to $-0.4$~W~m$^{-2}$. For comparison, the TOA energy budget ranges between $-0.2$ and 4.8 W~m$^{-2}$ in the preindustrial scenario with CMIP3 \cite[as shown in Fig.~2A in][]{lucariniragone2011}, and between $-3.16$ and 2.37 in CMIP5 \cite[see Table 2 in][]{lembo2019}. 
Moreover, the water-mass budget is null in all three PTB attractors. This is confirmed by negligible drifts over the last simulated 100~yr of the global mean ocean temperature and salinity, which are less than $1.5\cdot 10^{-2}$~$^\circ$C~century$^{-1}$ and $8\cdot 10^{-3}$~psu~century$^{-1}$ (in absolute value), respectively. 

\begin{figure}[t]
    \includegraphics[width=0.45\textwidth]{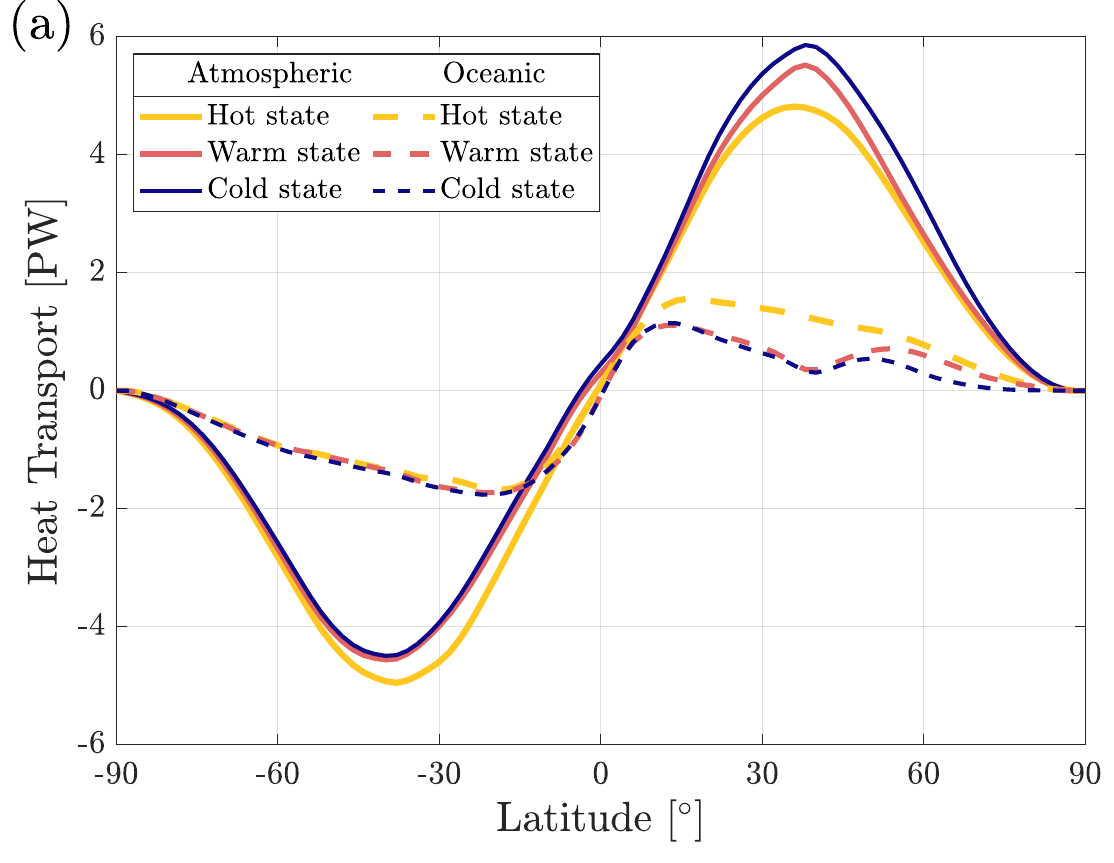}
    \includegraphics[width=0.462\textwidth]{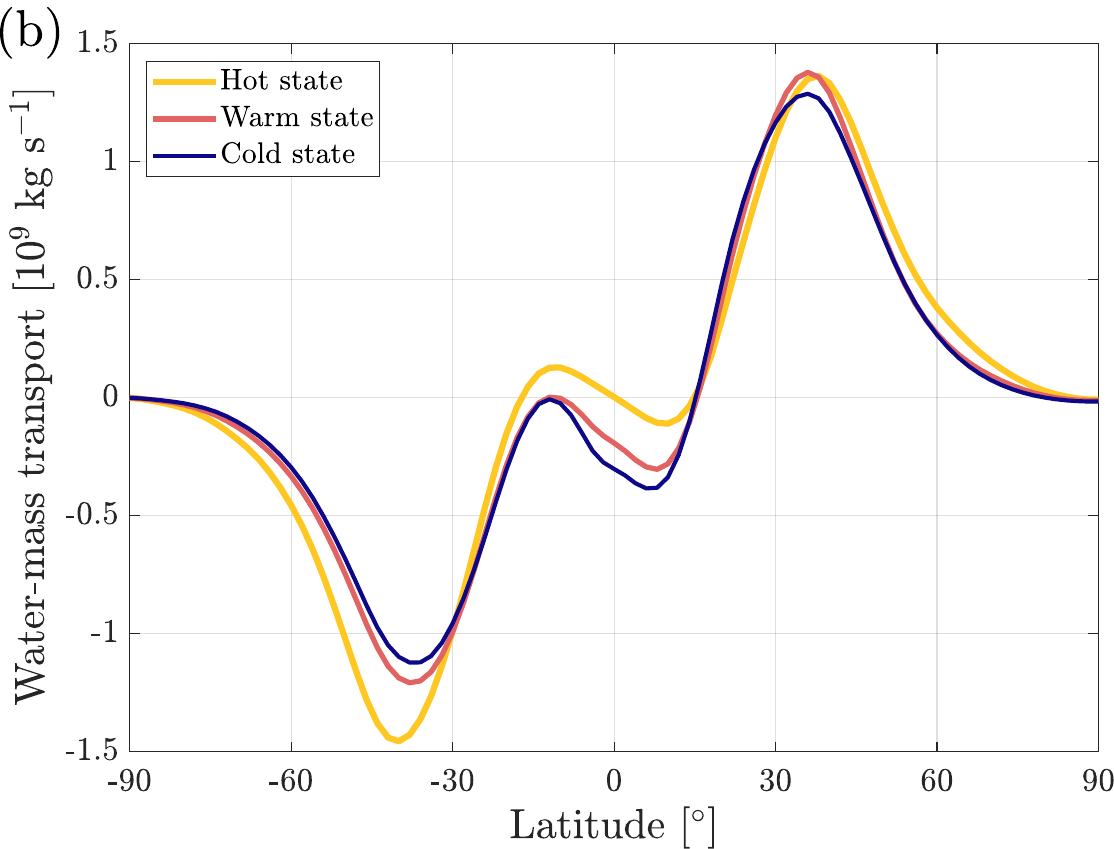}
    \caption{Climatological annual mean of (a) the northward meridional heat transport in the atmosphere (solid lines) and the ocean (dashed lines); (b) the northward meridional water mass transport in the atmosphere for the three attractors.}
    \label{fig:heatTransport}
\end{figure}

As reported in a coupled-aquaplanet configuration \cite{ragon2022}, the presence of sea ice increases the meridional surface air temperature gradient, 
and thus makes the atmospheric heat transport stronger than in an ice-free climate.
This is indeed what we observe in the northern hemisphere (NH), where the meridional temperature gradient is steeper in cold (28.5~$^\circ$C) than in warm (22.3~$^\circ$C) and hot (17.2~$^\circ$C) states. Thus, the intensity peak of the atmospheric heat transport is larger in the cold state in the NH (Fig.~\ref{fig:heatTransport}a). 
In contrast, the intensity peak of the oceanic heat transport is smaller in the cold state in the NH due to the structure of the overturning circulation (see below) and the presence of sea ice.  
In the southern hemisphere (SH), where sea ice is absent or negligible, the meridional temperature gradient is nearly equal in all states, implying a small variation in heat transport efficiency. 
In this case, other contributions to the total heat transport need to be accounted for to differentiate the states, as the latent heat and the associated water-mass transport (see Fig.~\ref{fig:heatTransport}b), which turns out to be more important in the hot state, where the whole hydrological cycle has greater intensity (i.e., more evaporation and precipitation, see Table~\ref{tab:globmean}). 

\begin{figure*}[ht!]
    \includegraphics[width=\textwidth]{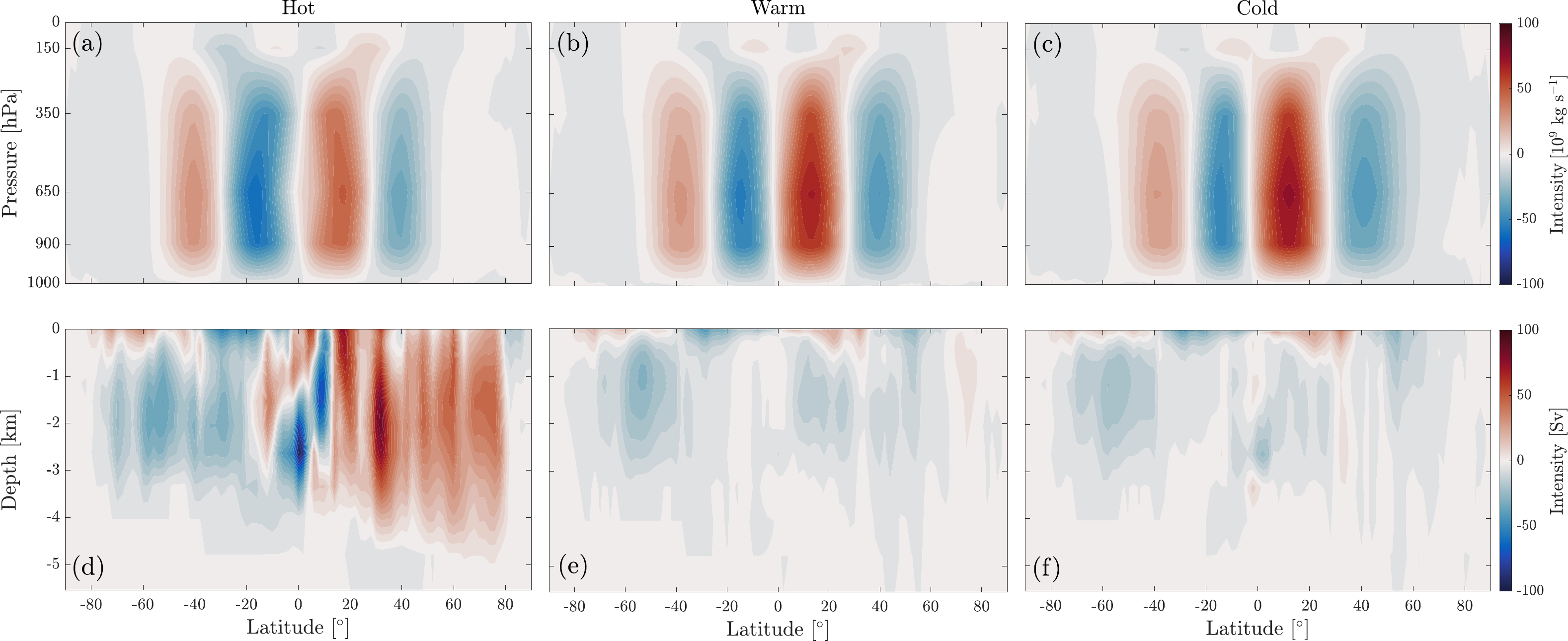}\hfill
    \caption{Annual mean overturning circulation in the atmosphere (a-c) and in the ocean (d-f) for the hot (a,d), warm (b,e) and cold (c,f) states. Color indicates streamfunction strength. Units are 10$^9$~kg~s$^{-1}$ for the atmosphere and sverdrups (1~Sv $=10^6$~m$^3$~s$^{-1}$) for the ocean. Positive and red ({\it resp.} negative and blue) correspond to clockwise ({\it resp.} anti-clockwise) circulation. Each color gradation corresponds to the transport of  $5\cdot 10^9$~kg~s$^{-1}$ (atmosphere) and 5~Sv (ocean).}
    \label{fig:overturning_atm_ocn}
\end{figure*}

We also observe an asymmetry between NH and SH in the mean annual atmospheric overturning circulation (Figs.~\ref{fig:overturning_atm_ocn}a-c). In the NH, the Hadley cell moves equatorward and becomes stronger in the cold state (see Fig.~\ref{fig:overturning_atm_ocn}a-c), leading to an increase of  $\sim3\cdot10^8$~kg~s$^{-1}$ in the equatorward moisture transport peak (Fig.~\ref{fig:heatTransport}b). Seasonal patterns are shown in Fig.~S1.
Note that when oceanic circulation is vigorous, like in the hot state (Fig.~\ref{fig:overturning_atm_ocn}d), the atmospheric circulation is weak (Fig.~\ref{fig:overturning_atm_ocn}a) compared to the other attractors, while a weak oceanic circulation leads to a strong atmospheric one, as seen in the cold state (Fig.~\ref{fig:overturning_atm_ocn}c and \ref{fig:overturning_atm_ocn}f). This is the well-known phenomenon denoted as Bjerknes compensation~\cite{bjerknes1964}.

\begin{figure*}[ht!]
    \includegraphics[width=\textwidth]{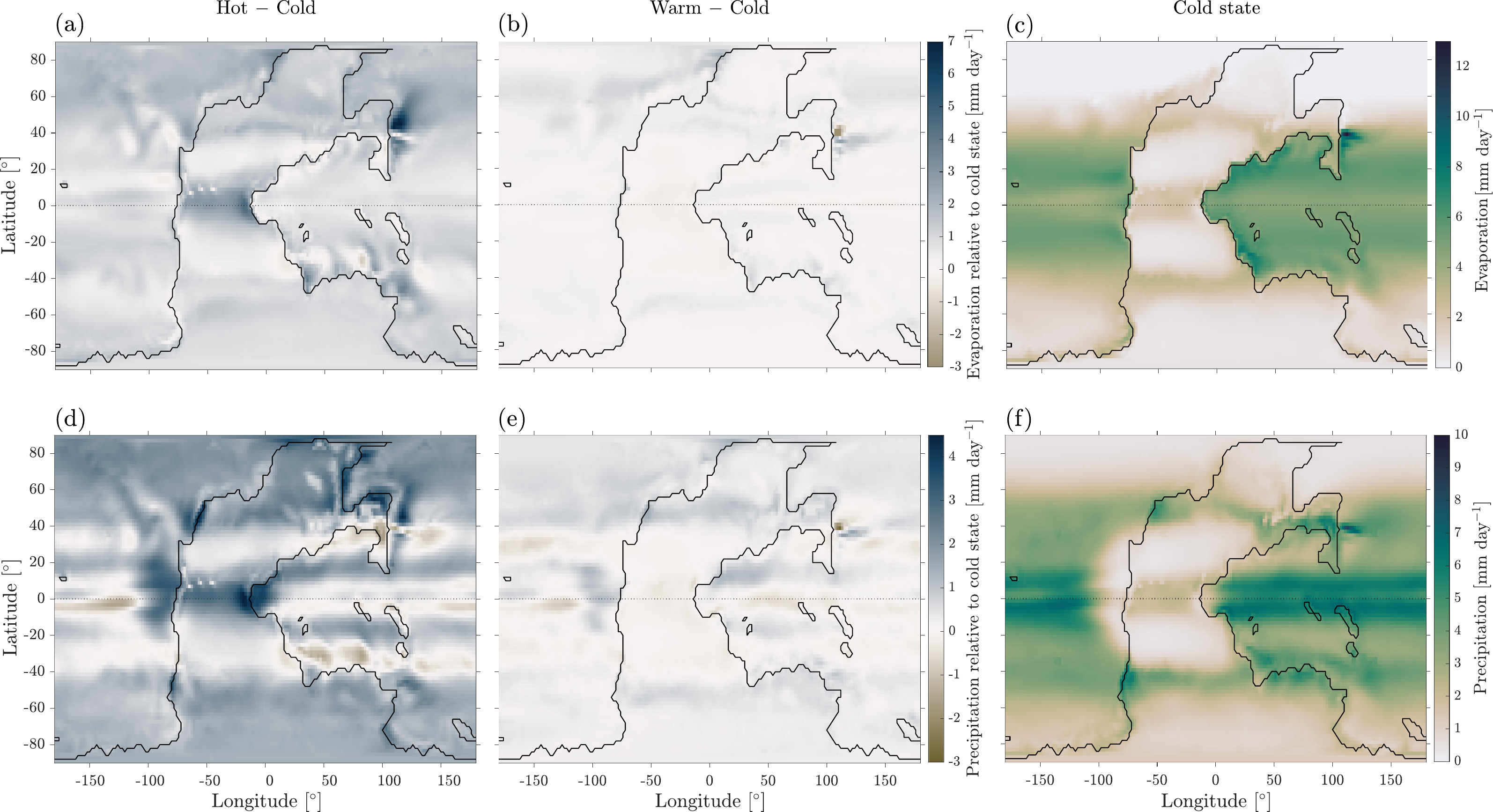}
    \caption{Evaporation (a-c) and precipitation (d-f) for the cold state (c,f) and differences between (a,d) hot and (b,e) warm states with respect to the cold state.}
    \label{fig:evap_precip}
\end{figure*}

Evaporation $E$ is maximal in oceanic tropical regions, while it is nearly zero over polar regions in the presence of sea ice, as seen in Fig.~\ref{fig:evap_precip}a-c. The highest rates of precipitation $P$ (Fig.~\ref{fig:evap_precip}d-f) are observed in equatorial regions, similar to pre\-sent-day climate. In the hot state, the amount of precipitation falling on land in the northern polar region remains significant. Dry continental areas, with low precipitation and evaporation, are present along the tropics in all three attractors.

The salinity distribution (sea surface maps are shown in Fig.~S2) 
is correlated with $E-P-R$, $R$ being the runoff. In the hot state, where sea ice is absent, the salinity distribution is symmetrical with respect to the Equator, as is the sea surface temperature (Fig.~\ref{fig:sat}), giving rise to a symmetrical configuration of the deep-water convection regions at the poles and of the oceanic overturning cells, with strong upwelling at the Equator (Fig.~\ref{fig:overturning_atm_ocn}d).
This symmetry disappears in the cold and warm states (Fig.~\ref{fig:overturning_atm_ocn}e and~\ref{fig:overturning_atm_ocn}f), where the absence of the clockwise overturning cell is associated with a drop in salinity and the presence of sea ice in the NH. In this case, the anti-clockwise cell extends across both hemispheres with a significantly reduced intensity compared to the hot state. 
Indeed, the intensity of the local maximum of the anti-clockwise overturning cell around 50~$^\circ$S decreases from hot (33~Sv) to warm (25~Sv) and cold (20~Sv), a trend consistent with~\cite{hulse2021}.
It is interesting to note that \cite{hulse2021} show similar patterns for the overturning cells at PTB for different values of atmospheric CO$_2$ content: the clockwise overturning cell becomes more intense as CO$_2$ increases, which agrees with the behavior observed in the hot state. 
Another interesting remark is that while the cold state has an average SAT larger than the pre\-sent-day value, its overturning circulation, which is constituted by a single anti-clockwise cell and thus is completely different from the present-day structure, allows for the formation of a large ice cap in the NH. Notably, the fact that the north polar region changes from upwelling in the cold state to downwelling in the hot state as a result of a different thermohaline circulation is in agreement with the interpretation of~\cite{amon2002} for the end-Permian collapse of the Permian Chert Event in the northwest margin of Pangea, due to the return to much warmer oceanic conditions.
Contrary to the thermohaline circulation, the overall structure of surface currents 
(Fig.~S3),
due to wind-driven circulation, is similar in the three attractors, although the subtropical gyres  are more symmetrical and intense in the two colder states. In terms of oceanic circulation, the warm and cold states display similar overall dynamics.

Finally, we have analysed the cloud cover in the three steady states. It is larger in the hot state than in the cold state in polar regions and on land 
(see Figs.~S4 and the resulting radiation at TOA in Fig.~S5).
The planetary albedo 
is 30\% in the cold state, 29\% in the warm state and 27\% in the hot state, which are less than or equal to the present-day  estimation~\cite{goode2001}. Thus, the energy absorbed into the atmosphere is generally greater than that for the present-day climate, and is larger in the hot than in the cold state.
At the same time, the atmospheric transmissivity of long-wave radiation is smaller in the hot (0.50) than in the cold state (0.57), meaning that more thermal radiation is trapped within the atmosphere in the hot state.
The combined effects of cloud feedback lead to more radiation entering and staying in the atmosphere in the hot state, with  cloud feedback becoming dominant, as also observed in the coupled-aquaplanet configuration analysed in~\cite{brunetti2019,zhuRose2022}. However, the cloud cover is a diagnostic variable in MITgcm and some feedbacks, like the negative one due to the reduction of ice relative to liquid content in clouds~\cite{bjordal2020}, are not included.






\subsection{Bifurcation diagram (BD) for varying atmospheric CO$_2$ content }
\label{section:BD}

\begin{figure}[t!]
\centering
\includegraphics[width=0.5\textwidth]{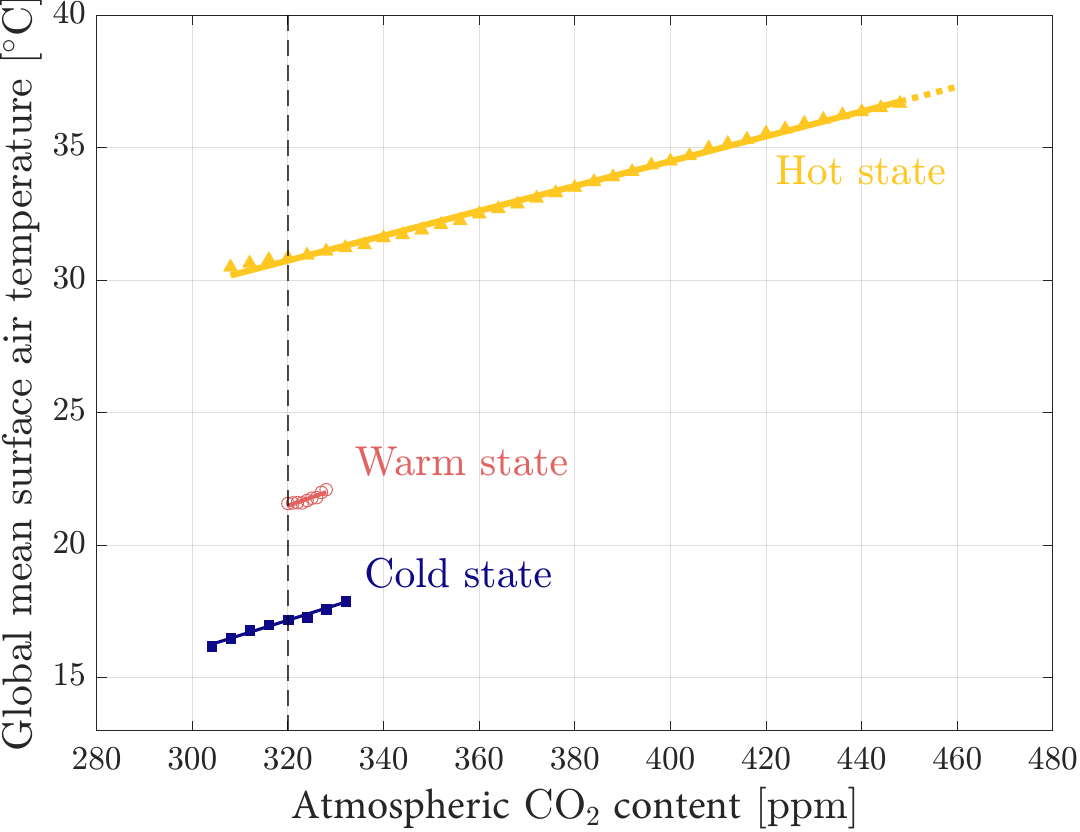}
    \caption{Bifurcation diagram in terms of the global mean surface air temperature as a function of the atmospheric CO$_2$ content. Markers correspond to averages over 100~yr for a given forcing value, error bars being within the marker size. The vertical dashed black line indicates the reference value pCO$_2 = 320$~ppm discussed in Section~\ref{section:attractors}.}
    \label{fig:BDsimplified}
\end{figure}

The three attractors described so far have been characterized for pCO$_2$~=~320~ppm (see Sect.~\ref{subsec:methodsBD}). By varying the atmospheric CO$_2$ content and thus modifying the radiative forcing, we identify the stability regions of each attractor and the associated tipping points through the BD construction 
(see Sec.~\ref{subsec:methodsBD} and Suppl. Sec.~2). 
The resulting stable branches are shown in Fig.~\ref{fig:BDsimplified}, where the steady values of the global mean SAT are displayed as a function of the CO$_2$ forcing. Climatic attractors are complex dynamical objects in a high-dimensional manifold~\cite{brunetti2019,Falasca2022}. The projection of their
invariant measure~\cite{Eckmann1985} on a given state variable is arbitrary~\cite{FarandaMessori2019,tel2020,brunetti2023}. Here, the projection is performed, as commonly done
in the literature, over the global mean SAT.

Tipping points are located at the endpoints of the stable branches, except for the right end of the stable range for the hot state, which could not be reached.
In fact, by increasing the forcing up to $\sim 500$~ppm along the stable branch of the hot state, the SAT increases to values larger than 38~$^\circ$C, where the model becomes numerically unstable. 
The limitation is due to the high temperature rather than to the pCO$_2$ value, since previous works for an aquaplanet~\cite{brunetti2023,moinat2024} and preliminary works for the present-day configuration manage to simulate larger pCO$_2$ values using the MITgcm, although for lower values of SAT.

Three attractors are found between 320 and 328~ppm (tri-stability). This region is limited by the extent of the warm state branch. Both the hot and cold branches extend beyond the warm one. Thus, through a B-tipping, it is possible to tip from the warm state to the cold or hot states, but it is impossible to reach the warm state from either the cold or hot states, which excludes the possibility of an hysteresis loop involving the warm state. 
The only way to attain the warm state would be to initially start in its basin of attraction, or to reach it through another tipping mechanism. However, if we assume that the size of the warm state basin of attraction is proportional to the branch length, catching the warm attractor would require very specific initial conditions. This suggests that the warm state may only have a low impact on the general dynamics of the climate system for the Permian-Triassic paleogeography.

A larger region of bi-stability from 304 to 332~ppm allows for the presence of a hysterisis loop between the hot and cold states. The temperature gap between these two attractors is on the order of 10~$^\circ$C. Thus, our numerical results suggest that tipping mechanisms can induce a shift from one attractor to another with climatic variations of this magnitude, while the carbon cycle was in a perturbed state as a consequence of the volcanic activity of the Siberian Traps~\cite{sun2012,romano2013,goudemand2019,leu2019,widmann2020}. It is interesting to note that there is a similar temperature gap of 10~$^\circ$C between previously published model outputs and geological records for the Permian-Triassic period~\cite{scotese2021,valdes2021}.  
Moreover, a temperature gap of this magnitude has been observed in climatic oscillations during the Early Triassic~\cite{widmann2020}, as mentioned in the Introduction, and at PTB~\cite{chen2016,gliwa2022}.

The solid lines in Fig.~\ref{fig:BDsimplified} are linear fits of the equilibrium temperature values on each branch. Their slopes $s$
are similar in the three attractors: $s_{\textrm{hot}} = 
0.0469\,(5)$~$^\circ$C~ppm$^{-1}$, $s_{\textrm{warm}} = 0.06\,(1)$~$^\circ$C~ppm$^{-1}$ and 
$s_{\textrm{cold}} = 0.057\,(3)$~$^\circ$C~ppm$^{-1}$. 
The equilibrium climate sensitivity (ECS) can be estimated from these slopes. The general formula, 
${\rm{ECS}} =  \Delta T/\log_2 (1+X)$, with $\Delta T = T-T_0$ and 
$X = ({\rm{CO}}_2-{\rm{CO}}_{2,0})/{\rm{CO}}_{2,0}$, can be linearised  for $X<1$ giving 
${\rm{ECS}} \sim s\, {\rm{ln}}(2) \,{\rm{CO}}_{2,0}$. This linearised formula, using ${\rm{CO}}_{2,0}=320$~ppm, gives ECS for the Permian-Triassic attractors ranging from 10.4~$^\circ$C to 13.3~$^\circ$C, much higher than
the ECS range for modern Earth, reported to be between 2~$^\circ$C and 5~$^\circ$C~\cite{IPCC2021_chap7}. 
The reason can be attributed to several factors: 
\begin{enumerate}

\item The ECS depends on the climatic state (deep time {\it vs} present day)~\cite{cp-11-1801-2015}; indeed, our preliminary simulations for the present-day climatic attractor with the same simulation setup of this study give ECS of around 5~$^\circ$C. 

\item There is a spread in modern-Earth ECS inferred by climate models, ranging between 1 and 8~$^\circ$C, which depends on models' complexity and the time scale considered~\cite{knutti2017}. Moreover, the ECS strongly depends on the cloud-feedback representation~\cite{bjordal2020}. 
As described in the Methods section, the atmospheric MITgcm module has a simplified parameterization for longwave radiation and cloud cover.  Despite these simplifications, the present-day climate can be reproduced with a MITgcm setup at the same horizontal resolution as used here, showing that the main feedback mechanisms are properly described at the main order, within the limitations of low-resolution GCMs. Indeed, the 
simulations described in \cite{brunettiverard2018} 
reproduce reasonably well the pre-industrial conditions, including  mean surface air temperature, sea ice extent in the Arctic and Antarctic regions, the structure and the maximal intensity of the overturning circulation,  as well as ocean and atmosphere heat transport. Furthermore, the TOA imbalance is quite low (-0.55~W m$^{-2}$), along with a surface energy imbalance of 0.04~W m$^{-2}$, ensuring a minimal temperature drift of only 0.009~K century$^{-1}$ ~\cite[Table~3 in][]{brunettiverard2018}.  
\end{enumerate}

The BD of Fig.~\ref{fig:BDsimplified} provides essential information on the climate system over the dynamical timescales considered in the numerical model. The longest timescale, corresponding to deep-ocean dynamics, is on the order of $10^3$~yr. Additional feedback mechanisms developing over similar or longer timescales need to be investigated using different numerical techniques, as discussed in the following.

\subsection{Vegetation cover on land}
\label{sec:veg} 

\begin{figure*}[ht!]
 \includegraphics[width=\textwidth]{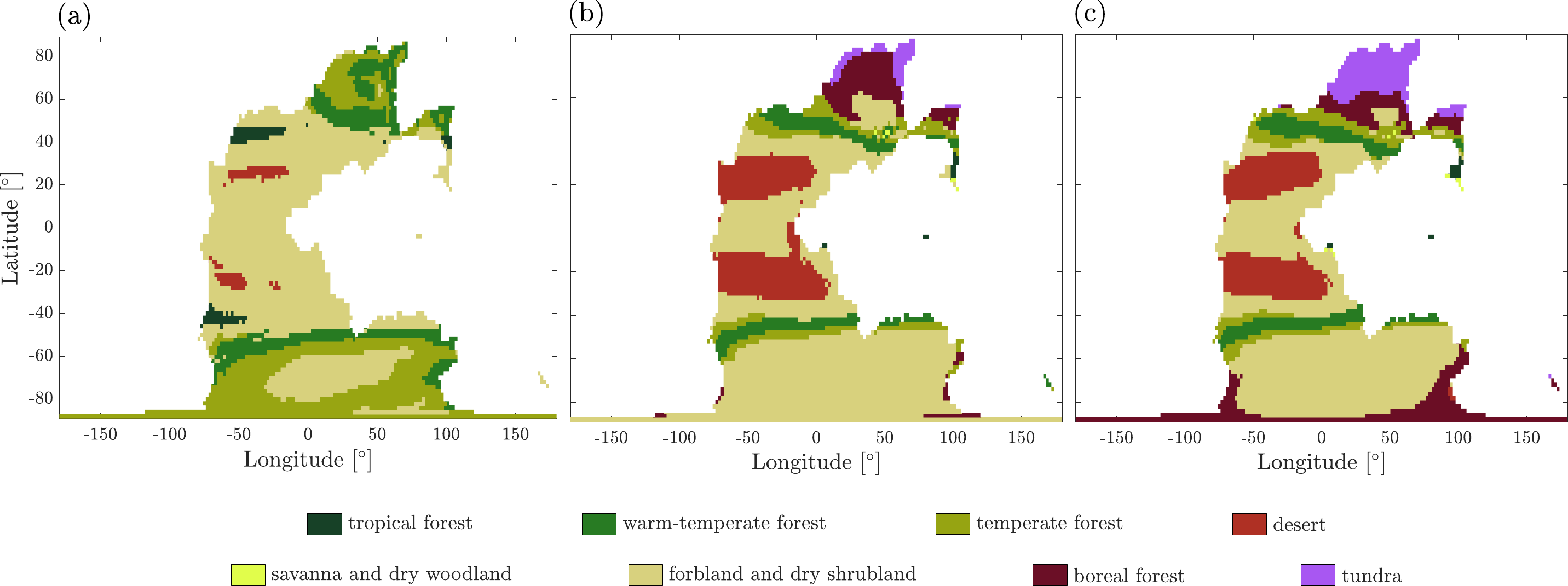}
    \caption{Vegetation cover represented through the major biomes defined in the legend from~\cite{harrison2003} for (a) hot, (b) warm, and (c) cold states with an atmospheric CO$_2$ content of 320~ppm. White area corresponds to ocean. Some small islands are not included in these maps (see Suppl. Sec.~3).}
    \label{fig:biomes}
\end{figure*}

Vegetation distribution provides an important feedback mechanism to the climate system, as it affects the albedo and evapotranspiration over land surfaces, thereby influencing the energy budget.
Moreover, the distribution and amount of terrestrial biomass play a crucial role in the global carbon cycle.
Here, we evaluate the long-term climate adjustment (up to 10$^4$~yr)
due to the vegetation cover at pCO$_2 = 320$~ppm using asynchronous coupling between MITgcm and the vegetation model BIOME4. 
We find that this coupling  does not change the number of attractors but 
shifts up their average SAT 
by 1.5~$^\circ$C in the hot state, 0.2~$^\circ$C in the warm state, and 0.8~$^\circ$C in the cold state. Fig.~\ref{fig:biomes} shows the resulting biome maps for the three attractors. For sake of clarity, biomes are grouped into major biomes~\cite[see classification in Table~3 of][]{harrison2003}. There is a quite good apparent fit between the biomes in some of the attractors and the lithologic indicators of climate developed by~\cite{boucot2013phanerozoic} around PTB.

In the cold state, desert areas are present in the tropics, surrounded by forbland. 
Moving poleward, the vegetation successively evolves to temperate/warm-temperate forests, to forbland again and to boreal forests. The forbland area is much larger in the SH, covering almost all longitudes between 60$^\circ$S and 80$^\circ$S. Tundra is present in the northern polar region, developing at the same latitudes as 
the sea ice extent. 
This suggests that ice sheets could develop instead of tundra if included in the simulation setup, thus albedo would be larger and the resulting climate even colder.
The warm state displays a similar trend, with the main differences occurring in the polar regions: in the North, boreal forests dominate over tundra, while in the South, forests are mainly replaced by forbland.
Finally, in the hot state, desertic areas nearly disappear, 
and forbland dominates in both tropical and subtropical regions. 
Polar regions are covered by successive slices of warm-temperate and temperate forests, 
as well as forbland in the South.

In order to understand why the desert has a smaller extent in the hot state than in the cold state, we provide seasonal SAT and precipitation in Figs.~S8 and S9, respectively, and their mean, minimum and maximum values in Fig.~S10.
Regions of zero precipitation remain consistent throughout the year in the cold state 
within 20 and 40$^\circ$, with mean SAT around 40~$^\circ$C. 
These regions correspond to desert in the cold state (Fig.~\ref{fig:biomes}).  
In contrast, regions of zero precipitation are much smaller and vary depending on the season in the hot state, 
despite reaching higher temperature than in the cold state, with a maximum around 60~$^\circ$C. 
Moreover, it is important to remember that SAT is not the only driver of desert conditions; soil temperature and soil water availability are crucial for the plant development as well~\cite{WAHID2007}. 
To estimate the amount of terrestrial biomass in each attractor, 
the 28 biomes in BIOME4 have been associated with ecosystem types and mean biomass densities (Table~S3).  
We find that the mean biomass density in the hot state is approximately 145~Mg~ha$^{-1}$ (corresponding to a total of $1.9\cdot 10^{17}$~mol of carbon globally), which is larger than 92~Mg~ha$^{-1}$ ($1.2\cdot 10^{17}$~mol\,C) in the cold state. 
While these values are about four times higher than the present-day ones~\cite{biomass2018} and may be overestimated, they clearly demonstrate the trend that the organic carbon stored in the vegetation cover is larger in the hot state than in colder climates.

\subsection{Air-sea carbon exchanges}
\label{section:carboncycle}

The carbon cycle describes how carbon is transferred between the atmosphere, hydrosphere, biosphere, cryosphere and lithosphere over different timescales. 
In our simulation setup, we can include how carbon flows between the atmosphere and ocean (Section~\ref{section:methodsCarbon}), but not among the other reservoirs.
Once equilibrium is reached, the carbon cycle acts to maintain a net zero flux at the interface between the atmosphere and ocean,  thus regulating the climate over $10^3$-$10^4$~yr. 

The advantage of including air-sea carbon exchanges is to obtain 
the carbon stored in the atmosphere and the ocean, 
and the spatial distribution of the air-sea CO$_2$ flux and other passive tracers for each attractor
(Sec.~\ref{section:methodsCarbon} and Suppl. Sec.~4). 
The effect is that both the atmospheric CO$_2$ content and SAT are only slightly changed at the equilibrium. Thus, the overall pattern of the BD is not affected (Fig.~S11).

As mentioned above, our simulation setup does not take into account carbon fluxes between lithosphere, biosphere and cryosphere, such as weathering, biomass in terrestrial plants, or carbon stored in the permafrost.
Thus, the carbon in the atmosphere and the ocean that we have estimated is affected by these simplifications. Since the precipitation 
(Figs.~\ref{fig:evap_precip}d-f) and the mean biomass density 
(Section~\ref{sec:veg}) are much larger in the hot state than in the cold one, neglecting weathering and biomass fluxes has a larger impact on hot than cold climates. Interestingly, the difference in biomass carbon, which we estimate to be on the order of $10^{17}$~mol in Section~\ref{sec:veg}, lies in the same range as the difference between the
amounts of the carbon content in the cold state ($3.0 \cdot 10^{18}$~mol) and in the hot state ($2.8 \cdot 10^{18}$~mol). Since vegetation and air-sea carbon exchanges act on the same time scale, the biomass can indeed significantly contribute to this difference.


\section{Concluding remarks}
\label{section:conclusions}

The climate is a nonlinear system that can display multistability, reflecting the fact that there is no 
unique way to redistribute energy when several feedbacks are active on the same time scale~\cite{GhilLucarini2020}. 
BDs are an important tool to understand the backbone dynamical structure of the system,  
which can include several attractors, particularly to identify stability and multistability regions, the position of tipping points, aand the amplitude of forcing and 
internal variability needed for tipping~\cite{brunetti2023}.
In the case of the Permian-Triassic paleogeography produced by PANALESIS, we found three attractors,  
which we characterised in detail, and compared to other numerical simulations from the literature. 
We constructed the BD for varying pCO$_2$ (Fig.~\ref{fig:BDsimplified}), where bi-stable and tri-stable regions can be identified. 


In the considered setup, 
feedback mechanisms occurring on time scales comparable to or longer than the relaxation time of the deep-ocean circulation are excluded and need to be accounted for using alternative numerical techniques. 
We restarted simulations at selected positions in the BD 
by including relevant long-term feedbacks, namely the evolution of the vegetation distribution (via asynchronous coupling between MITgcm and BIOME4) and the exchanges of carbon between the atmosphere and ocean (through the activation of additional packages in MITgcm). 
These methodologies enabled us to incorporate long-term adjustments in the BD and rectify the position of its stable branches, resulting in only minor modifications.
Similar techniques could be employed to account for other relevant long-term mechanisms, such as the evolution of ice sheets, which are currently not implemented in our setup.

While absolute values may depend on the specific model used, our main finding is that multistability can exist in a general circulation model with a realistic paleogeography.
By exhibiting a temperature gap of approximately $10~^\circ$C between the hot and cold states, the BD near the PTB opens up the possibility of explaning some climatic variations observed in geological records of the Early Triassic, notably near the Smithian-Spathian boundary~\cite{widmann2020}. These variations could be attributed to the hysteresis loop existing between these two attractors (B-tipping), increased internal variability due to the biological pump or volcanism (N or S-tipping), or a forcing mechanism which varies over time with a critical rate (R-tipping).
Additionally, a temperature difference of 10~$^\circ$C between climatic states could explain the disparity between surface temperatures reconstructed by geological records~\cite{scotese2021} and those obtained by numerical models~\cite{valdes2021} 
near the PTB.

Constructing BDs with the same feedback mechanisms but using other numerical models would be necessary to identify robust characteristics and  reduce artefacts of models and boundary conditions. 
In particular, including sea ice dynamics or different numerical implementations of thermodynamic sea ice~\cite{Lewis2007,voigt2012,esd-15-215-2024}, as well as considering a mixed-layer ocean or a fully dynamical one~\cite{poulsen2001,pohl2014}, could alter the number of steady states, and reveal the source of potential biases. 
The construction of BDs represents a promising method for investigating the dynamics of the climate system both in deep-time and under present-day forcing setups. 

\section{Methods}
\label{section:methods}

\subsection{Model description}
\label{subsec:model}

Numerical simulations are performed using the MIT general circulation model~\cite[MITgcm, version c67f,][]{marshall_finite-volume_1997,marshall_hydrostatic_1997,adcroft_implementation_2004} 
in a coupled atmosphere-ocean-thermodynamic sea ice-land configuration 
with a fixed Permian-Triassic paleogeography. 

The atmospheric module is based on SPEEDY \cite{molteni_atmospheric_2003}, which provides a simplified description of the convective scheme,  large-scale condensation,  vertical diffusion,  surface fluxes and  radiative transfer with diagnostic clouds. In particular, in the parameterization scheme for the longwave radiation, the infrared spectrum is partitioned into four regions: 1) the `infrared window' between 8.5 and 11~$\mu$m; 2) the band of strong absorption by CO$_2$ around 15~$\mu$m; 3) the aggregation of regions with weak/moderate absorption by water vapour; 4) the aggregation of regions with strong absorption by water vapour. 
The coarse vertical resolution (5 layers, with the top one representing stratosphere) together with the simplified assumptions in the parameterization schemes allow for one order of magnitude less CPU time than a state-of-the-art atmospheric GCM at the same horizontal resolution, making SPEEDY suitable for studies on millennial time scales.
Taking
into account the limitations due to the coarse vertical
resolution and the simplified representation of
physical processes, SPEEDY provides a rather realistic
simulation of the atmospheric flow~\cite{molteni_atmospheric_2003}. 

The oceanic dynamic component is like in state-of-the-art climate models, with 28 vertical levels. It accounts for tracer diffusion~\cite{redi1982} and advection of geostrophic eddies~\cite{GentandMcWilliams1990}
through the Gent and McWilliams scheme, as well as vertical mixing \cite{large1994}. 
Sea ice is described by a purely thermodynamical three-layer module \cite{winton2000}, and land by a two-layer module.

We use a cubed-sphere grid with $32\times 32$ points per face (CS32), which corresponds to 2.8$^{\circ}$ horizontal spatial resolution. Cloud albedo is varying as a function of latitude~\cite{Kucharski2013,ragon2022}
and frictional heating is re-injected into the system to guarantee the approximate closure of the energy budget at Top-of-Atmosphere (TOA)~\cite{brunettiverard2018,brunetti2019,ragon2022,zhuRose2022}. 
Earth's rotation period is set to 22.2~h \cite{arbab2009} to account for Permian-Triassic conditions.
The solar constant is estimated to be 1336~W~m$^{-2}$~\cite{gough1981,foster2017} .

\begin{figure}[t]
\centering
\includegraphics[width=0.7\textwidth]{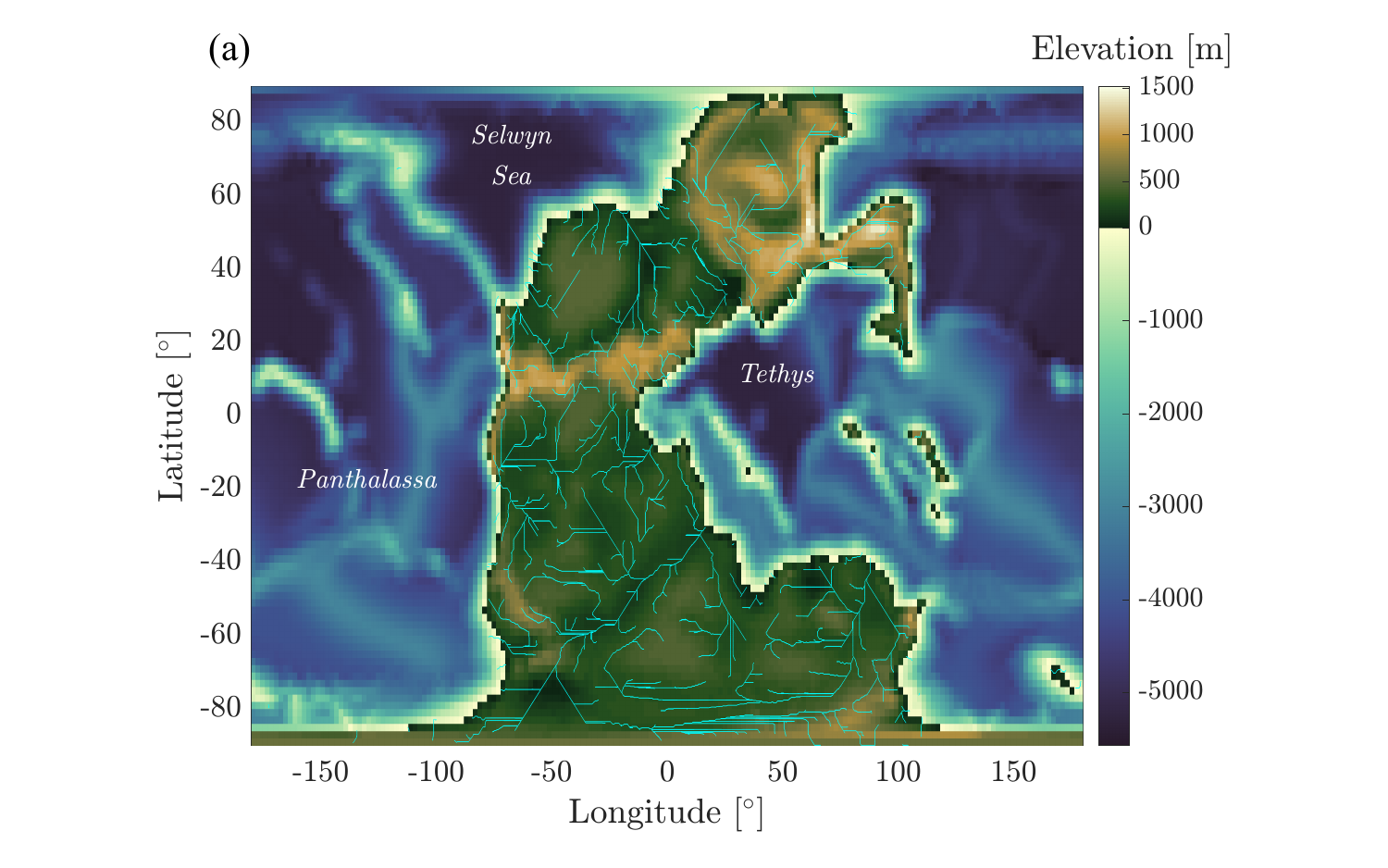}  
\includegraphics[width=0.48\textwidth]{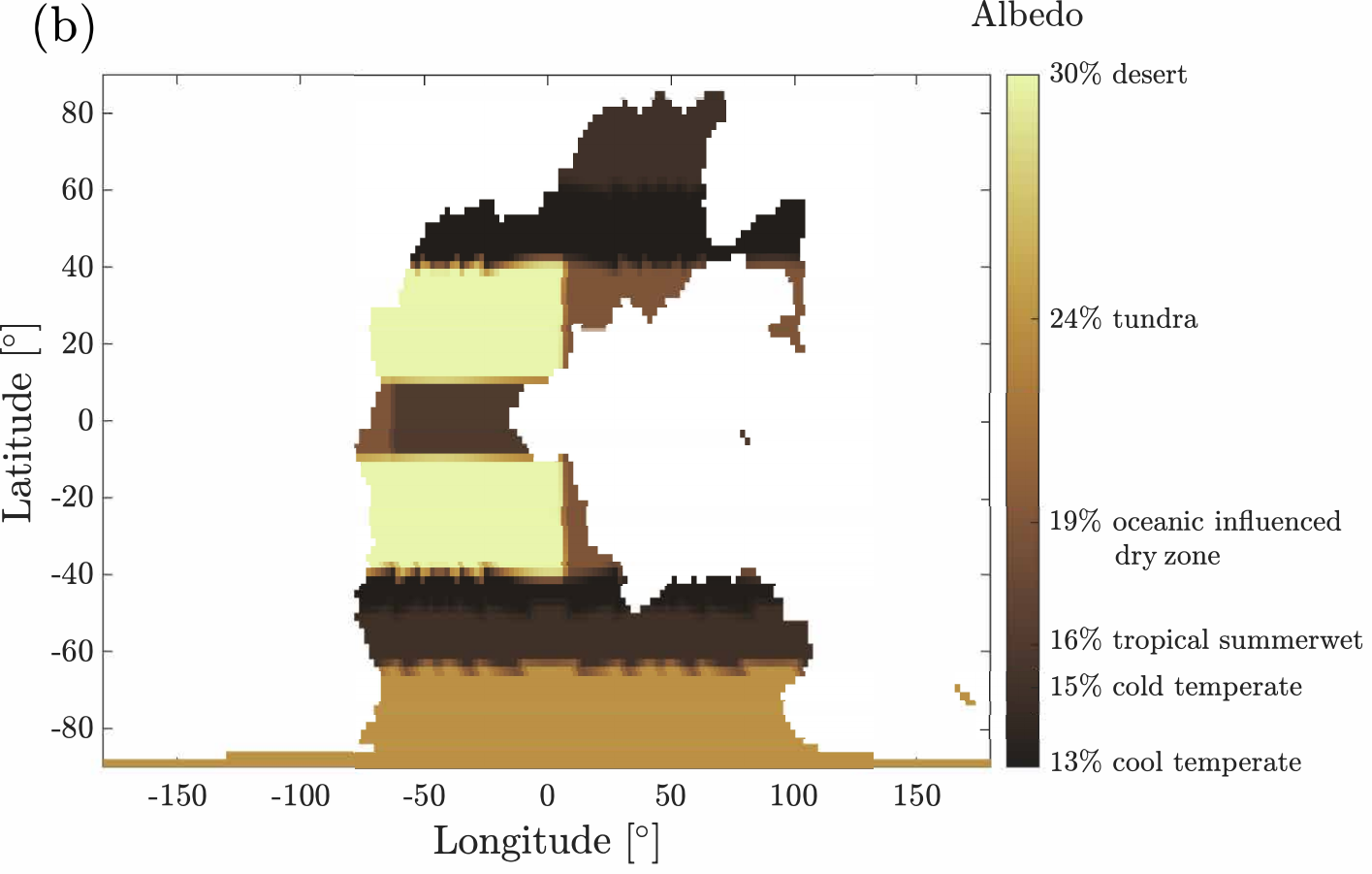}
\includegraphics[width=0.48\textwidth]{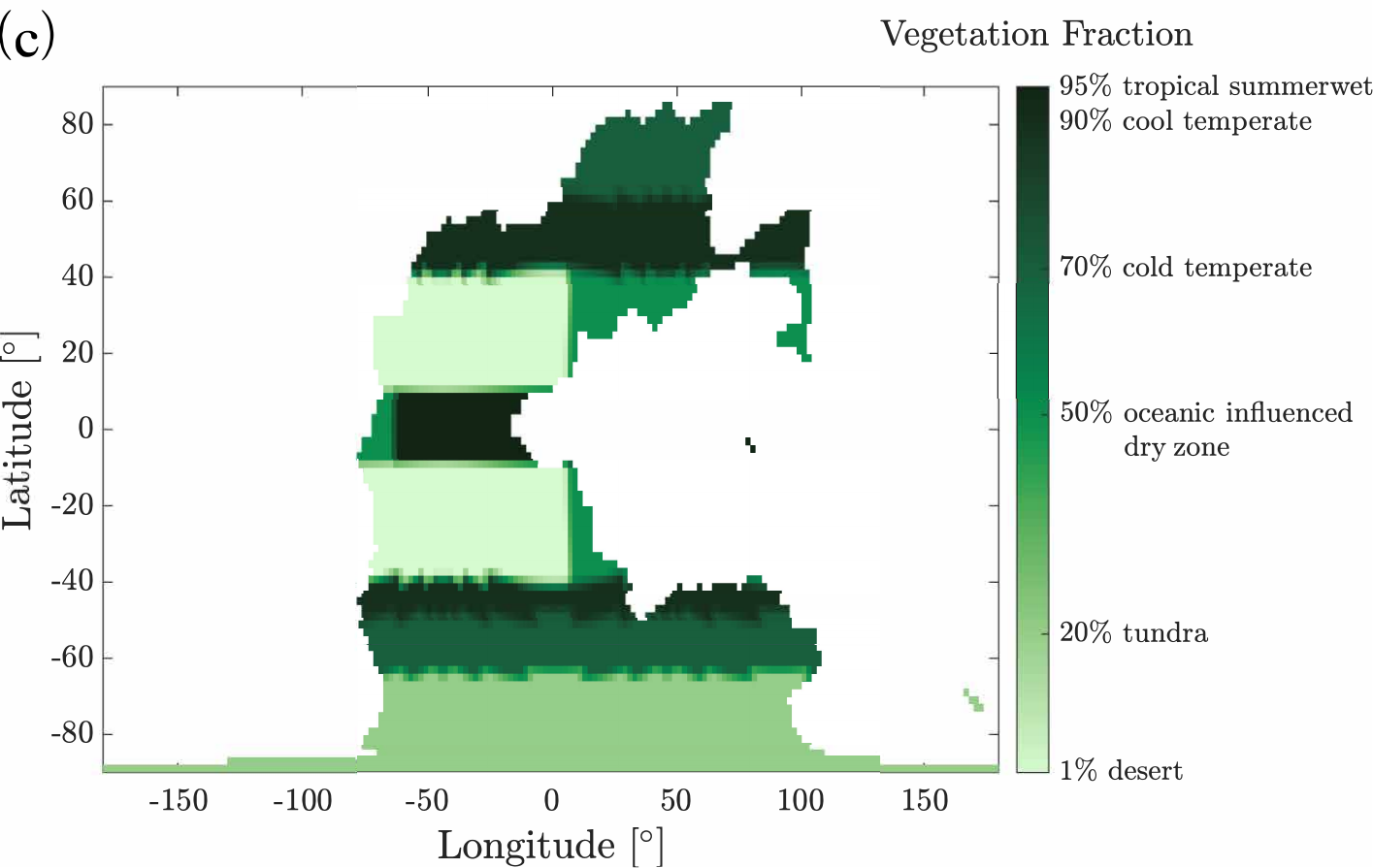}
    \caption{(a) Paleogeography of the Permian-Triassic reconstruction. Rivers are represented in cyan and are used for the runoff map. 
    Initial vegetation cover and albedo (b), as well as vegetation fraction (c) used in the climate simulations with fixed vegetation cover. }
    \label{fig:map}
\end{figure}

Our simulations rely on the paleogeography produced by PANALESIS~\cite{verard2019panalesis,verard2021},
a global plate tectonic model providing maps every 10~Ma from 888~Ma (Tonian) to the present. The global reconstruction of the Permian-Triassic paleogeography features a large continental mass, Pangea, extending from the southern to the northern polar regions, surrounded by a wide oceanic realm, Panthalassa. Two more oceans are present, Tethys at equatorial latitudes and the Selwyn Sea in the northern polar region. 
Since narrow seaways result in unrealistic ice accumulation and numerical instabilities of the climate model, the original PANALESIS map was adapted to the model horizontal resolution, specifically by enlarging seaways narrower than a few pixels or closing the smallest ones, as well as closed epicontinental seas or lakes. 
The resulting topography used in the simulations is displayed in Fig.~\ref{fig:map}a.
Land and oceanic packages are linked via a runoff map, which is obtained by defining drainage basins and the corresponding main rivers, as represented
in Fig.~\ref{fig:map}a, so that each land point is associated to an ocean point that corresponds to the revelant river mouth.
The initial zonal distribution of the vegetation cover is derived from Figs.~6.D-E-F of \cite{Rees2002}.
The corresponding values of bare soil albedos and vegetation fractions, shown in Fig.~\ref{fig:map}, are used as boundary conditions in the climate simulations. Note that the evapotranspiration coefficient is not taken into account in the moisture flux. 

\subsection{Search for attractors and construction of the bifurcation diagram}
\label{subsec:methodsBD}

Estimates of atmospheric CO$_2$ content at the PTB and the subsequent climate oscillations during the Early Triassic are affected by huge uncertainties \cite{retallack2001,Joachimski2022,shen2022}. Reported values range from less than 100~ppm in the Late Permian to more than 1700~ppm in the Early Triassic, according to the range of 95\% confidence interval from Fig.~1 of~\cite{foster2017}. Our modelling strategy is thus the following: first, 
we set the atmospheric pCO$_2$ to an intermediate value (320~ppm, the default value in MITgcm; moreover,  
we set the MITgcm parameter {\tt{aim\_select\_pCO2}} $= 1$ to exclude feedback with the ocean). Using the method described in \cite{brunetti2019}, we perform dozens of simulations by varying the initial conditions and let the system relax towards a climatic attractor. This is done either by 1) using different mean values of oceanic temperature; 2) transiently varying some parameters of internal processes, such as the relative humidity threshold for low cloud formation or the atmospheric CO$_2$ content, in order to perturb the radiative budget at TOA and obtain a different climate trajectory. 
Afterwards, the parameters are restored to their original values. 
Second, starting from the attractors obtained at the previous step, we construct the stable branches of the BD by exploring a large range of pCO$_2$, using Method~II from \cite{brunetti2023}. More precisely, we slightly increase or decrease the forcing by $\Delta \textrm{pCO}_2$~=~2--4~ppm
at regular temporal intervals of $\Delta \textrm{t}$~=~100~yr. This allows us to construct the full BD and identify, in particular, the position of B-tipping.

Since also temperature reconstructions are affected by huge uncertainties~\cite{scotese2021}, and climatic oscillations observed in the stratigraphic record at a temporal resolution of $10^5$~yr reveal the presence of cooling phases with a temperature gap of around 10$^\circ$C~\cite{widmann2020}, our simulations cannot be tuned to any particular value around PTB. We have tested in present-day simulations that by tuning some parameters (like the threshold in relative humidity for low clouds formation) the attractor branches shift as a whole, either up or down in SAT. Thus, the dynamical structure of the climate attractors, as represented in the BD, is much more robust than their precise position in phase space. 

\subsection{Evolution of the vegetation cover}
\label{section:methodsVegetation}

After the identification of the climatic attractors, we use asynchronous coupling between the MITgcm and BIOME4 models to estimate the vegetation cover corresponding to each attractor. 

BIOME4 is a coupled carbon and water-flux model~\cite{HaxeltinePrentice1996,kaplan2001,kaplan2003} that is driven by long-term averages of monthly mean temperature, sunshine and precipitation to identify 28 different biomes for a given value of atmospheric CO$_2$ content. In addition, BIOME4 requires information related to soil texture and soil depth. For this, we use the global average of typical present-day values provided by BIOME4 itself, namely water holding capacity (110.1~mm~m$^{-1}$ for the first soil layer and 137.6~mm~m$^{-1}$ for the second one) and percolation rate (5.2~mm~hr$^{-1}$ for both layers).
BIOME4 is based on the concept of Plant Funcional Type (PFT), rather than taxonomic grouping \cite{kaplan2001}. 
Although the definition of PFTs is constructed on present-day plants, this is the only available framework that can be applied to plants at different deep-time geological periods, assuming that their functionality was similar to that of present-day plants. 
Within this framework, plants with certain fundamental characteristics (i.~e., growth form, phenology, rooting depth) are grouped together and included in BIOME4, allowing for  the study of their distribution at the  global scale. BIOME4 has already been applied to study the Jurassic climate \cite{SELLWOOD2008},
the Middle Pliocene \cite[3.6-2.6 Ma;][]{salzmann2008}, and the Last Glacial Maximum \cite{harrison2003}. 

The procedure for coupling MITgcm and BIOME4 is detailed in Suppl. Sec.~3. 
\subsection{Inclusion of the air-sea carbon flux}
\label{section:methodsCarbon}

Exchanges of carbon between the ocean and the atmosphere are particularly strong in regions of deep-water formation, upwelling or boundary currents. Since these processes regulate the climate over time scales of the order of 10$^3$-10$^4$ yr~\cite{zhuRose2022}, they can affect the BD based on deep-water dynamics.
Including dynamical carbon exchanges between the ocean and the atmosphere (through the activation of the GCHEM, DIC and PTRACERS modules in MITgcm)~\cite{follows2006}
doubles the CPU time compared to the configuration without these modules, thus preventing the use of this option from the start for the construction of BDs. 
The less consuming option we chose is to activate air-sea carbon exchanges after the convergence of the simulations towards the attractors, at specific positions on each stable branch, namely, at pCO$_2$ = 320~ppm and at the edges of the stable branches.
The new configuration now includes the evolution of five passive tracers: dissolved inorganic carbon (DIC), dissolved organic phosphorus (DOP), alkalinity, phosphate and oxygen. 
For this study, we use version c67j of MITgcm, which includes a new implementation of the pH solver that allows for  modelling  more extreme ocean temperature conditions than  present-day ones~\cite{munhoven2013}.

The numerical procedure is divided into two steps:
\begin{enumerate}
    \item we use the reference profiles of the five passive tracers for  present-day Earth provided by MITgcm as initial conditions at 320~ppm.
    We then compute the distributions of tracers which are consistent with an atmospheric CO$_2$ content of 320~ppm by keeping this value constant (this is done by setting the model parameters {\tt{aim\_select\_pCO2}} = 1 and {\tt{dic\_int1}} = 1).
    At the edges of the stable branches, the distributions of tracers obtained for pCO$_2$ = 320~ppm are used as initial conditions to reduce spin-up time. We stop the simulations of this first step when the annual CO$_2$ flux at the ocean surface, averaged over the last 100~yr, is lower than 0.1~ppm yr$^{-1}$.
    
    \item The tracer profiles obtained at the previous step are then used to restart the model with dynamical air-sea carbon exchanges.
    In this case, by setting the model parameters {\tt{aim\_select\_pCO2}} = 3 and {\tt{dic\_int1}} = 3, the DIC module provides the evolution of the air-sea CO$_2$ flux, of the five biochemical tracers, and the global carbon content in the atmosphere and the ocean. Note that, by setting initial tracers concentrations to values different from the distributions obtained at step 1, for example directly using the reference MITgcm profiles, a spurious amount of carbon would enter the global carbon reservoir, creating a perturbation that could push the system into another attractor.
\end{enumerate}

%

\begin{acknowledgements}
We are grateful to Jan-Henrik Malles for running some of the simulations in the first stage of this work. We thank all the Sinergia project members (PaleoC4, \url{https://www.unige.ch/paleoc4/}), Emmanuel Castella,  St\'ephane Goyette and Laure Moinat for very useful discussions. C.~R. and M.~B. thank the MITgcm-support mailing list for valuable advice on the code. The simulations were performed on the Baobab and Yggdrasil clusters at the University of Geneva. We acknowledge the financial support from the Swiss National Science Foundation (Sinergia Project No. CRSII5\_180253).
\end{acknowledgements}

\smallskip\noindent
{\bf Author contribution: } {M.~B. planned the study, C.~V. provided the paleogeographic reconstruction, C.~V. and C.~R. obtained the runoff map, C.~R. and M.~B. performed the climate simulations, C.~R. produced the final numerical results and figures, all the authors were involved in the analysis and discussion of the results. C.~R. and M.~B. wrote the manuscript, and all the authors discussed and edited the final paper.} 

\smallskip\noindent
{\bf Competing interests:} {The authors declare no competing interests.} 

\smallskip\noindent
{\bf Code data availability:} 
{The data supporting the findings of this study were generated
by the MIT general circulation model, which is openly
available on GitHub (\url{http://mitgcm.org/, https://github.com/MITgcm/MITgcm}, versions c67f and c67j), and by BIOME4 (\url{https://github.com/jedokaplan/BIOME4}).} 


\noindent
{\bf Supplementary Information} The online version contains supplementary material available on the Scientific Reports web page.

\printbibliography  
         

\end{document}